# Effect of droplet-induced fluid motion on the interphase synthesis of nanoparticles in a microfluidic reactor

Vivekananda Bal, Rajdip Bandyopadhyaya*

Department of Chemical Engineering, Indian Institute of Technology Bombay,

Powai, Mumbai 400076, India.

* Corresponding author (rajdip@che.iitb.ac.in)

**Abstract**

Droplet-based interphase synthesis provides means to produce nanoparticles with low polydispersity by controlling mass transport through droplet dynamics. An experimentally validated model, based on coupled computational fluid dynamics with population balance equation is proposed. The model incorporates: (i) hydrodynamics and thermal-transport of droplet-laden flow, (ii) spatial variation of species concentration with interfacial mass transfer and (iii) a tertiary phase within droplets to capture evolution of particle size distribution (PSD). We conclude that, mixing both inside and outside the droplets controls the PSD (arithmetic mean diameter ($D_{10}$), ~ 50 nm).Consequently, as spherical-droplet diameter increases from 71% to 96% of the channel diameter (812 μm), coefficient of variation (CV=standard deviation/mean diameter) of PSD decreases from 33% to 21.3%, implying less polydisperse nanoparticles, due to altered mixing. With non-spherical drops, CV is further reduced to only 11%, which is significantly lower compared to that of 40% in single-phase micro-reactor, suggesting benefit of former.

# 1 Introduction

In the past few decades, there have been many experimental research works on nanoparticle synthesis in a segmented flow microreactor.[1-10] However, none of these explicitly looked at the effect of droplet induced mixing dynamics on the dynamics of particle evolution therein. Though there has been a prior attempt[1] to analyze mixing dynamics based on a simple diffusion effect, it is preliminary in nature and does not offer in-depth analysis. Even the effect of droplet size and shape on the mass transport and resulting effect on particle evolution still remains unexplored. Understanding this mass transport dynamics would help not only the experimentalists and theoreticians involved in nanoparticle synthesis, but also offer insights in other areas[1] like extraction, separation, purification, protein crystallization etc. Therefore, in the present work, we concentrate on mass transport limited nanoparticle formation dynamics in a microreactor.

In case of segmented flow synthesis, there has been a plenty of attention towards the liquid-liquid dispersion based segmented flow[1] (liquid-liquid slug flow, liquid-liquid droplet flow) based microreactor synthesis. This is because of numerous advantages over laminar flow single phase synthesis - namely reduced reactor fouling,[2] reduced (Taylor-Aris) dispersion and residence time distribution effects - resulting in a narrow particle size distribution,[3-5] augmented heat transport[6-7] and enhanced mixing due to internal circulation of liquid.[7-9] Even though, almost all of the drawbacks of laminar flow, single phase synthesis is overcome in gas-liquid segmented flow reactor[7] (gas-liquid slug flow, gas-liquid bubble flow), it is seen to suffer severely from drawbacks of producing a broad particle size distribution.[2-3,11-19]

Though there have been a lot of theoretical investigations on the single phase stirred tank and single phase microreactor synthesis processes,[20-39] there is hardly any model for segmented flow microreactor synthesis of nanoparticles, because of the non-linear coupling between fluid dynamics inside and outside the drop with temporal evolution of particles. Earlier theoretical works in a segmented flow reactor solely dealt with either heat transfer characteristics[7] or hydrodynamics of the flow.[40-53] It was found that, in presence of either drops or bubbles, heat transfer rate is augmented by 4-5 times than that in the single phase.[7, 45-46] In order to model the hydrodynamics of the flow, their model used volume of fluid (VOF) algorithm with a continuum surface force model, proposed by Brackbill et al.[54] Later on, Fischer et al.[6] further extended the previous studies, by introducing droplets containing nanoparticles, by improving previous models to include the effect of actual physical curvature of interface and solving VOF, along with nanoparticle continuity equation. Their study revealed that, radial heat transfer rate is strongly dictated by the interfacial tension between two liquids and overall thermal transport is further augmented by the presence of nanoparticles inside the droplets, though the increment is very low (3-5%).

Principal objective of this paper is to theoretically investigate the droplet based interphase synthesis of nanoparticles, wherein reagents are initially present in different phases, as it provides better control over reaction[55] to form growth units, compared to droplet based single phase synthesis. In order to achieve this, the whole problem has been decoupled into two parts: in the first part, we numerically study the hydrodynamics and forced-convective thermal transport in droplet-laden flow, using VOF multiphase method, with an interface tracking scheme. In the second part, we study the particle formation dynamics by solving the PBE (population balance equation), including all particle events namely reaction, nucleation, diffusion growth, coagulation and Ostwald ripening (OR), and coupling the effect of already established velocity and temperature field as obtained from the first part. Simulation results have been compared with the experimental results reported by Zukas and Gupta[55] for ZnO synthesis. Finally, we present a theoretical analysis on the effect of droplet size, shape on the mean size and size distributions of particles and compare with that of corresponding batch and laminar flow single phase reactor. Modeling interphase synthesis is far from trivial and to the best of our knowledge, there is no prior work on the multiphase synthesis of nanoparticles, including all intricate details, wherein reagents are initially present in two separate phases. There are some works on the multiphase synthesis of nanoparticles, where reagents are present in separate phases, such as microemulsion synthesis in a batch reactor,[56-57] but in all these cases, spatial variation of reactants has not been considered. Instead, their models consider a lumped rate of species transport. In the microfluidic synthesis, broader objective has been to achieve particles with a smaller size and a narrow size distribution. Present work is a leap forward, as it explains the physics of nanoparticle formation in a droplet based set up to explicitly help experimentalists to understand the physics of particle formation therein. Apart from this, the present work also provides insight on the operating regime for a segmented flow based microfluidic reactor.

## 2 Model formulation

It is difficult to track the particle formation inside the droplet and simultaneously solve for hydrodynamics of droplet laden flow. Hence, the whole problem has been decoupled into two parts: In the first part of the work, hydrodynamics and energy transport of the droplet laden flow have been simulated in the whole microfluidic reactor (see the schematic of reactor in supplementary material section S1) and in the second part, particle formation has been modeled in a unit cell[considering a single droplet and surrounding continuous fluid phase (figure 1)], including the effect of velocity and temperature distribution from the first part. Concept of the unit cell has been conceived such that, it's length is equivalent to center to center distance between two consecutive droplets and it is periodic in nature. In the present work, classical volume-of-fluid multiphase (VOF) algorithm has been employed to model the motion of droplets along with the continuous phase.

### 2.1 Continuity equation and momentum balance equation

Continuity equation[58, 37-38] for phase $q$ is written as

$$\frac{\partial}{\partial t}(\alpha^q \rho^q) + \nabla \cdot (\alpha^q \rho^q \bar{u}^q) = \left( \dot{m}^{ps} - \dot{m}^{sp} \right) + S^q \tag{1}$$

where, $q$ represents either primary phase ($p$) or secondary phase ($s$). Here, primary phase is the continuous fluid phase (i.e carrier fluid 1-Octanol) and secondary phase is the droplet phase (i.e water). $\bar{u}^q$ is $q^{th}$ phase linear velocity vector (m/s). $\alpha$ is the secondary phase volume fraction, $\rho^q$ is the $q$ phase density, $\dot{m}^{ps}$ is the mass transfer rate (kg/m$^3$s) from phase $p$ to $s$ and $\dot{m}^{sp}$ mass transfer rate (kg/m$^3$s) from phase $s$ to $p$, $S^q$ is the source term.

The momentum balance equation[58] for combined phase is given as

$$\frac{\partial}{\partial t}(\rho_m \bar{u}) + \nabla \cdot (\rho_m \bar{u}\bar{u}) = -\nabla P_r + \nabla \cdot \bar{\bar{\tau}} + \rho_m \bar{g} + F_{sv} \tag{2}$$

where, $\rho_m$ is the combined phase density (kg/m³), dependent on each phase volume fraction. $g$ is the gravity. $\bar{u}$ is the linear velocity vector, $P_r$ represents the pressure (N/m²), $\bar{\bar{\tau}}$ is the shear stress tensor, $F_{sv}$ represents the volumetric force (N/m³). Surface force ($F_{sa}$) at the interface (N/m²), which has been modeled by Brackbill et al.,[54] is given as follows

$$F_{sa} = -\sigma_{ll}\kappa_c \hat{n} + \nabla_s \sigma_{ll} \qquad (3)$$

where, $\sigma_{ll}$ and $\kappa_c$ are the surface tension (N/m) and curvature (m⁻¹) of freely deformable liquid-liquid interface, $\nabla_s$ is the surface tangential gradient operator and $\hat{n}$ is the unit normal vector at the liquid-liquid interface. Here, surface curvature is calculated from divergence of local gradients of volume fraction of $q^{th}$ phase at the interface. In order to incorporate the surface force into the momentum balance equation (2), surface force ($F_{sa}$) has been converted into volume force ($F_{sv}$). (see supplementary material section S7).

Initial and boundary conditions for solving the above equations are

IC: $\bar{u} = 0$ (inside the domain)
BC: $\bar{u} = 0$ at the walls
BC: $\bar{u}.\hat{n} = u$ (inlet) and $\bar{\bar{\tau}}.\hat{n} = 0$ (outlet)

## 2.2 Energy balance equation

In case of energy balance equation too, single energy balance equation is solved and simulation results are shared by all the phases. The corresponding energy balance equation[58, 38] is given by

$$\frac{\partial(\rho_m E_m)}{\partial t} + \nabla.[\bar{u}(\rho_m E_m + P_r)] = \nabla.(k_{eff} \nabla T_m) + S_h \qquad (4)$$

where, $k_{eff}$ is the effective thermal conductivity (W/mK), $S_h$ is the volumetric energy source term, which is zero in this case. $E_m$ and $T_m$ are the mass-averaged energy (Joule) and temperature (K).e.g. the mass averaged energy is given as

$$E_m = \frac{\sum \alpha^q \rho^q E^q}{\sum \alpha^q \rho^q} \quad (5)$$

where, $E^q$ is the $q^{th}$ phase energy. $E$ is related to the temperature through the relationship given as $E_m = C_{p,m}(T_m - T_{ref})$. where, $T_{ref}$ is the reference temperature, $C_{p,m}$ is the mass average specific heat of mixture phase. Room temperature, 298 K, has been taken as the reference temperature. Specific heats of water and n-octanol are 4.2 J/gm °C and 308 J/mole K, respectively [6, 59, 71].

IC: $T_m = T_i$ (initial temperature inside the reactor)
BC1: $T_m = T_w$ (constant temperature at walls of reactor)
BC2: $T_m = T_i$ (inlet of microreactor)
BC3: $-k_{eff} \nabla T_m = 0$ (outlet of microreactor)

## 2.3 Species balance equation

Once velocity profile is obtained, species transport equation is solved inside a unit cell [consisted of droplet phase (secondary phase; *s*) and continuous fluid phase (primary phase; *p*) surrounding the droplet], as shown in figure 1. Initially, both reactants are in separate phase. NaOH diffuses inside the droplet and reaction starts. Once particles are formed a new phase called tertiary phase (*t*) is introduced into the system. The corresponding species transport equation for NaOH is given as

$$\frac{\partial}{\partial t}\left(\alpha^q \rho^q Y_{NaOH}^q\right) + \nabla \cdot \left(\alpha^q \rho^q \bar{u}^q Y_{NaOH}^q\right) = -\nabla \cdot \alpha^q \hat{J}_{NaOH}^q + \alpha^q R_{NaOH}^q + \sum \left(\dot{m}_{NaOH}^{ps} - \dot{m}_{NaOH}^{sp}\right) \quad (6)$$

where, $Y_{NaOH}^q$ is mass fraction of NaOH in phase $q$ (either *p* or *s* phase), $\hat{J}_{NaOH}^q$ is the diffusion flux of NaOH in phase $q$, $R_{NaOH}^q$ is the loss or gain of NaOH in phase $q$, $\dot{m}_{NaOH}^{ps}$ is the transfer of NaOH from phase *p* to *s* and $\dot{m}_{NaOH}^{sp}$ is the mass transfer rate of species from phase *s* to *p*. Species balance equation for NaOH is solved in both primary and secondary phases as NaOH diffuses inside the droplet for the reaction to happen.

Initial and boundary conditions for solving the above equation are as follows

Outside droplet (primary phase, $p$; 1-octanol):

IC: $Y^p_{NaOH} = Y_{NaOH,0}$

BC1: $-\overline{N}_{NaOH}.\hat{n} = 0$ (insulation at walls of reactor)

BC2: $Y^p_{NaOH} = Y_{NaOH,0}$ (at boundary 1)

BC3: $-D_m \alpha^p \rho^p \nabla Y^p_{NaOH} = 0$ (boundary 2)

BC4: $-\overline{N}^p_{NaOH}.\hat{n} = k_m \left( \dfrac{\alpha^p \rho^p Y^p_{NaOH}}{K_{ow}} - \alpha^s \rho^s Y^s_{NaOH} \right)$, Flux continuity at droplet interface where, $\overline{N}_{NaOH}$ is the mass flux of NaOH, $k_m$ is the mass transfer coefficient of NaOH, which has been taken as the diffusivity divided by the thickness of diffusion layer (according to film theory). $K_{ow}$ is the partition coefficient of NaOH between primary and secondary phase (see supplementary material section S8 for more details). $D_m$ is the diffusivity of NaOH.

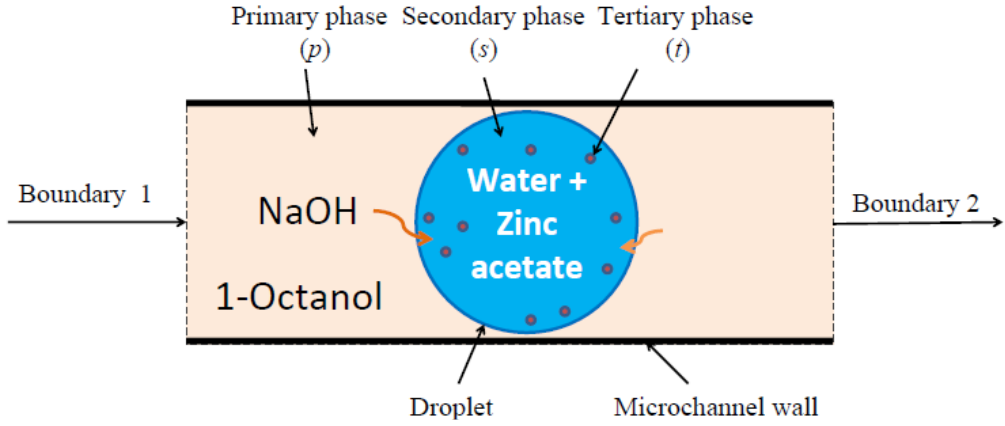

**Figure 1:** Unit cell for particle formation modeling

The corresponding species transport equation for zinc acetate is given as

$$\frac{\partial}{\partial t}\left(\alpha^s \rho^s Y^s_{Zn(OAc)_2}\right) + \nabla.\left(\alpha^s \rho^s \overline{u}^s Y^s_{Zn(OAc)_2}\right) = -\nabla.\alpha^s \hat{J}^s_{Zn(OAc)_2} + \alpha^s R^s_{iZn(OAc)_2} \quad (7)$$

where, $Y^s_{Zn(OAc)_2}$ is mass fraction of zinc acetate in phase $s$, $\hat{J}^s_{Zn(OAc)_2}$ is the diffusion flux of zinc acetate in phase $s$ and $R^s_{Zn(OAc)_2}$ is the loss or gain of zinc acetate in phase $s$ due to reaction. Species balance equation for Zn(OAc)$_2$ is solved in secondary phase ($s$; water inside droplet) only.

Inside droplet (secondary phase, $s$; water)

IC: $Y^s_{Zn(OAc)_2} = Y_{Zn(OH)_2,0}$

BC: $-\overline{N}_{Zn(OAc)_2} \cdot \hat{n} = 0$ (droplet interface), where $\overline{N}_{Zn(OAc)_2}$ is the mass flux of $Zn(OAc)_2$.

The corresponding species transport equation for Zinc oxide is given as

$$\frac{\partial}{\partial t}\left(\alpha^s \rho^s Y^s_{ZnO}\right) + \nabla \cdot \left(\alpha^s \rho^s \overline{u}^s Y^s_{ZnO}\right) = -\nabla \cdot \alpha^s \hat{J}^s_{ZnO} + \alpha^s R^s_{ZnO} + \sum\left(\dot{m}^{ts}_{ZnO} - \dot{m}^{st}_{ZnO}\right) \quad (8)$$

where, $Y^s_{ZnO}$ is mass fraction of zinc acetate in phase *s*, $\hat{J}^s_{ZnO}$ is the diffusion flux of zinc acetate in phase *s*, $R^s_{ZnO}$ is the loss or gain of zinc acetate in phase *s* due to reaction, $\dot{m}^{ts}_{ZnO}$ is the transfer of zinc acetate from phase *t* (tertiary) to *s* and $\dot{m}^{st}_{ZnO}$ is the mass transfer rate of zinc acetate from phase *s* to *t*. Species balance equation for ZnO is solved in secondary phase only.

Inside droplet (secondary phase, *s*; water)

IC: $Y^s_{ZnO} = 0$

BC: $-\overline{N}_{ZnO} \cdot \hat{n} = 0$ (droplet interface), where $\overline{N}_{ZnO}$ is the mass flux of ZnO.

## 2.4 Modeling particle formation

Classical population balance equation (PBE) equation has been solved in the tertiary phase (solid phase) inside the droplet in order to model particle formation dynamics. The PBE for number density is given as (Ramakrishna;[59] Randolph and Larson[60])

$$\frac{\partial n(v,x,t)}{\partial t} + \nabla \cdot [\overline{u} n(v,x,t)] + \nabla_v \cdot [G_v n(v,x,t)] = \frac{1}{2}\int_0^v q_B(v-v',v')n(x,v-v',t)n(x,v',t)dv' \\ - \int_0^\infty q_B(v,v')n(x,v,t)n(x,v',t)dv' \quad (9)$$

where *v* and *v'* are the particle volumes, *n(v, x, t)* is the number density function (number/m$^6$) of particles, $\nabla$ represents the operator in spatial co-ordinate, $\nabla_v$ represents the operator in particle volume co-ordinate, $q_B$ is the Brownian coagulation kernel (m$^3$/s). $G_v$ is the volumetric growth rate (m$^3$/s) of particles. In this analysis, we are studying the variation of arithmetic mean ($D_{10}$) and standard deviation of the distribution. Therefore, we are solving full form of PBE instead of solving the first and second moments, even though the later can reduce the complexity of the problem.

**2.4.1 Nucleation**

Classicalnucleation theory for homogeneous nucleation[61-62] has been employed to model nucleation rate ($n_0$; number/m³s), which is given as:

$$n_0 = A \, \exp\left[-\frac{16\pi\sigma^3 v_m^2}{3(k_B T)^3 (\ln S)^2}\right] \quad (10)$$

where, $A$, $\sigma$ and $v_m$ are pre-exponential factor (number/m³s), interfacial tension (N/m) and molecular volume (m³) of ZnO. $S = c_{ZnO}/c_{ZnO,s}$ is the supersaturation level of ZnO, where, $c_{ZnO}$ and $c_{ZnO,s}$ are bulk concentration and solubility (kg/m³) of ZnO in phase $s$. (see supplementary material section S2.1, S2.2 and S2.3 for detailed discussion)

**2.4.2 Diffusion growth**

Once nuclei are formed, diffusion limited growth starts by adding molecules of ZnO from the bulk of the medium to the solid ZnO particle and the corresponding growth rate expression is obtained by solid phase mass balance. This is given as

$$\frac{dr}{dt} = \frac{D_m v_m N_A c_{ZnO}}{r} \quad (11)$$

where, $N_A$, $D_m$, $r$ are the Avogadro's number, molecular diffusion coefficient of ZnO in liquid medium and radius of particle, respectively. (see supplementary material section S3)

**2.4.3 Coagulation kernel**

In this case, as the diameter of ZnO particle is very small, flow-induced coagulation of particles is essentially not taking place. Therefore, Brownian coagulation mostly dominates the coagulation process. The classical Brownian coagulation kernel ($q_B$) proposed by Smoluchowski (Smoluchowski;[63] Bal and Bandyopadhyaya[37-38]) is given by

$$q_B = \beta \frac{2k_B T}{3\mu} \left( \frac{1}{v_i^{\frac{1}{3}}} + \frac{1}{v_j^{\frac{1}{3}}} \right) \left( v_i^{\frac{1}{3}} + v_j^{\frac{1}{3}} \right) \tag{12}$$

where, $\mu$ is the viscosity (kg/ms) of the water in the droplet [6]. Viscosity of the solution is assumed to be not changing as the solution is dilute. $v_i$ and $v_j$ represent volumes (m³) of colliding particles and $\beta$ is the coagulation efficiency.

### 2.4.4 Ostwald ripening (OR)

The rate of Ostwald ripening (dissolution growth of particles) is obtained by solid phase mass balance, considering diffusion limited process. The corresponding rate expression[64-65] is given as below

$$\frac{dr}{dt} = \frac{D_m v_m N_A (c_{ZnO} - c_r)}{r} \tag{13}$$

where, $c_r$ is the concentration (kg/m³) of ZnO molecules at the surface of the particles. This is a function of particle radius, $r$, and the corresponding expression is given by Gibbs-Thomson equation.[64-67] as follows

$$c_r = c_{ZnO,s} \exp\left( \frac{2\sigma v_m N_A}{RT} \frac{1}{r} \right) \tag{14}$$

$c_{ZnO,s}$ is the solubility of ZnO. (see supplementary material section S4 for more discussion).

### 3 Experimental method

In order to validate our model, simulation results have been compared with experimental data published in literature by Zukas and Gupta[55] for droplet based interphase synthesis of ZnO nanoparticles. A detailed description of the experimental procedure, as proposed by Zukas and Gupta,[55] has been discussed in supplementary material section S1. To briefly describe, experimental procedure consists of a total flow rate of 0.165 ml/min through a microchannel of diameter 0.812 mm. Flow rate is kept constant throughout the experiment and channel length was varied to obtain the desired residence time.

## 4 Method of computation and validation of code

### 4.1 Method of computation

Numerical simulation of model equations was done using well known finite volume method based commercial software Fluent 14.5. Tetrahedral elements were used to mesh the whole computation domain. Linear (Energy balance, species balance, volume fraction equations) and non-linear (momentum balance equation) unsteady state governing PDEs were integrated over the control volumes, with piecewise linear approximation for conductive flux terms and second-order upwind spatial discretization scheme for convective terms (Patankar[68])in order to generate a set of linear and non-linear ODEs, respectively. Similarly, population balance equation (PBE) was integrated over control volumes, with second-order upwind scheme for convective flux term, to obtain a set of linear and non-linear PDEs. This is further discretized in particle volume domain, with the method proposed by Hounslow et al.[69] to obtain a set of linear and non-linear ODEs. Set of ODEs thus obtained was temporally discretized by integration in time domain with a second-order implicit temporal discretization scheme to obtain a set of linear and non-linear algebraic equations. These algebraic equations were solved using the iterative method with an under-relaxation scheme. Pressure-velocity coupling was achieved with phase coupled SIMPLEC algorithm. Codes for nucleation, diffusion growth, Brownian coagulation and OR were written in C++ and used as user defined functions in fluent.

Mesh convergence test was performed and it was confirmed that, solution does not improve on further increase in number of control volumes from a total number of 563920 volume elements. Throughout the whole computation, time step-size wasaltered between $10^{-8}$-0.01 sec, in order to ensure independence of final results with respect to time-step. Quantitative convergence of the solution was achieved, when scaled absolute residual reached below $10^{-5}$ for all variables, except for energy. In case of energy balance, quantitative convergence was achieved when scaled absolute residuals reached a value below $10^{-6}$. Mass balance check was also performed and it was found that systems' total mass remains conserved with < 1.1% error in mass balance.

### 4.2 Validation of finite volume based code

We validated our finite volume based Fluent code for four different limiting cases. These are: (i) validation of Fluent code for fluid mechanics part of droplet laden-flow, (ii) validation of coupled fluid dynamics and energy transport code in case of droplet laden-flow, (iii) validation of PBE and species balance equation code in absence of fluid flow and (iv) validation of PBE and species balance equation code in presence of single-phase fluid flow. These have been shown and discussed in the supplementary material section S9.

## 5 Results and Discussion

As mentioned earlier, in the first part of the model-development work, we investigate only the hydrodynamics and heat transfer characteristics (without any reaction), in a droplet-laden multiphase flow inside a microfluidic reactor. Specifically, we study the influence of introducing (pure immiscible) droplets of different size, shape etc. on energy and momentum transport. Subsequently, in the second part of the model-development, an experimentally validated simulation of inter-phase synthesis of nanoparticles in a droplet has been presented.

### 5.1 Effect of drop size on flow field and thermal transport

In the previous analysis by Fischer et al.,[6] it has been shown that, introduction of droplets in the single phase flow increases the heat transfer performance. They also found that for systems with low Reynolds number flow, temperature dependence of surface tension (Marangoni effect) can be seen only in the entrance region. Therefore, in the present analysis, we ignore the Marangoni effect as both drop Reynolds number ($Re_d$ = 2.4) and bulk Reynolds number ($Re_b$= 0.48) are low. So, we first investigate the effect of introducing droplets of different sizes (without nanoparticles), keeping all other parameters same as that in the experiments of Zukas and Gupta.[55] For convenience of computation, we have assumed that flow has reached the fully developed form, before entering the microreactor. Droplets are introduced into the microreactor of internal diameter of 0.812 mm in an interval of ~1s, as calculated from Zukas and Gupta.[55] The corresponding hydrodynamics of droplet-laden flow has been shown in figure 2 for mean inlet velocity of 5.3 mm/s. Here, it is observed that, in presence of droplets, stretching of central fluid layers with respect to the boundary layers near the tube-wall (as in the parabolic velocity-profile of a single-phase) is reduced, since flow in the continuous phase is hindered in presence of droplets.

As drop size increases, this effect is more pronounced and complete segmentation of flow takes place as shown in figure 2a. It is also observed that, there is circulation of fluid both inside the drop and in the segmented region between two droplets, shown by streamlines in figure 2a, as obtained by both Fischer et al.[6] and Urbant et al.[7] In contrast to this, for smaller drops, recirculation of fluid is completely absent, as shown by streamlines in figure 2b.

In order to investigate the effect of larger droplets on hydrodynamics, non-spherical drops of larger volume have been introduced into the system, keeping the droplet center to center distance and flow rate the same as in the previous case. The corresponding streamlines are shown in figure 2c. It is observed that, circulation of fluid is more pronounced both inside and outside the droplets, due to a larger blockage effect, as compared to that observed in case of corresponding spherical drops (Fig. 2a).

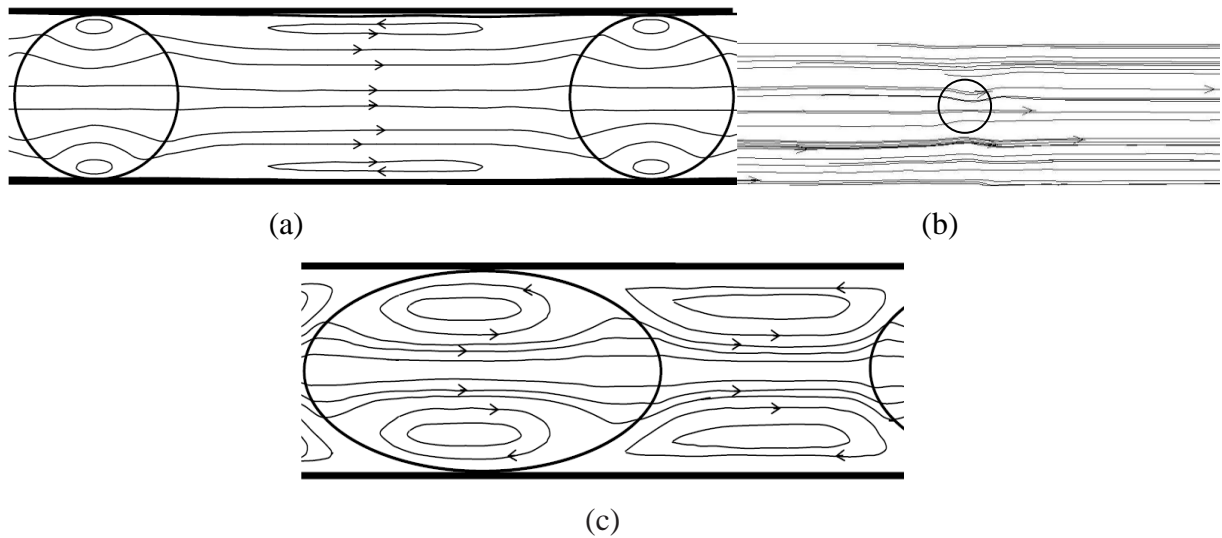

(a)                                  (b)

(c)

**Figure 2:** Effect of introducing droplets of different sizes ($d$=droplet diameter) into a microreactor of diameter, $D_c$=0.812 mm (Zukas and Gupta,[55]), $T$=25 °C. Parameters: $\rho$ = 998 kg/m$^3$ and $\mu$= 0.0089 kg/ms for drop (water), and $\rho$=824 kg/m$^3$ and $\mu$= 0.00736 kg/ms for bulk liquid (n-octanol) [71]. Streamlines for (a) large spherical drop ($d$=0.78 mm), (b) small spherical drop ($d$= 0.37 mm), and (c) non-spherical drop of large volume (minor axis length=0.78 mm and major axis length=1.45 mm). Center to center distance between two drops = 2.34 mm

Next, to investigate the effect of hydrodynamics on energy transport in the present system, we compare in figure 3a, the aforementioned cases of both small and large drops, and find out the temperature distribution at different times, once it enters the reactor. In case of larger drops (figure 3a), it is found that the thermal boundary layer is strongly affected due to the strong recirculation region near the wall, as shown in fig 3a. As an immediate effect of this, temperature inside the drop is found to increase at a faster rate, due to increase in energy transport inside the droplet in the region close to the micro-channel wall. The latter occurs because of the dragging of the hot fluid from the thermal boundary layer to the drop interior. Towards the center of the drop, circulation induced energy transport is reduced and consequently energy transport mostly takes place due to conduction. In the external region surrounding the front of the drop, energy transport is faster, because of the circulation-induced transport of hot fluid from thermal boundary layer (referring to the direction of the streamlines in the convection-loops in fig. 2a or 2c), whereas in the external region at back of the drop, relatively cold fluid is transported towards it, from the bulk region and therefore, temperature rises slowly (fig. 3a). In the case of smaller drops, enhancement of energy transport is not well pronounced (fig. 3b), because of the insignificant change in hydrodynamic condition (fig. 2b) from that in case of a single phase (Supplementary material S6). Consequently, temperature inside the drop increases at a slower rate, as shown in figure 3b. In fact, it is found that,temperature gradient at the center of the droplet is very less, in case of larger drops (fig. 3a), whereas in case of smaller drops there still exits a large temperature gradient, in the same time period of 0.51 s (fig. 3b).

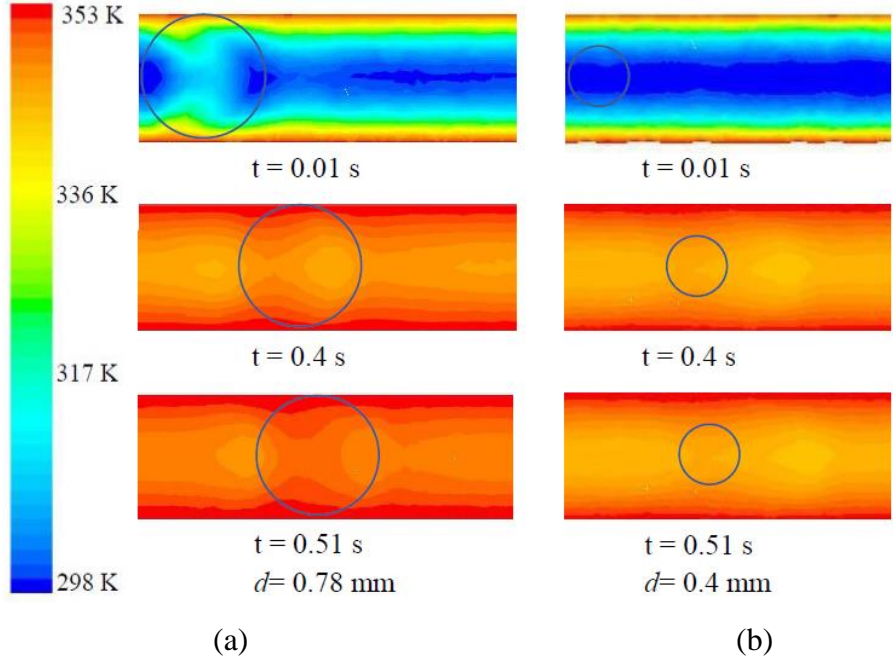

**Figure 3:** Effect of droplet hydrodynamics on energy transport in a microreactor of diameter $D_c=$ 0.812 mm (Zukas and Gupta[55]). Parameters: $\rho$ = 998 kg/m$^3$ and $\mu$= 0.0013 kg/ms for drop (water), and $\rho$=824 kg/m$^3$ and $\mu$= 0.00736 kg/ms for bulk liquid (n-octanol) at $T$=25 °C. Viscosity and density values at different temperatures are calculated from Bhattarcharjee and Roy [71] as both varies almost linearly with temperature. Streamlines for (a) large drop ($d$=0.78 mm) and (b) small drop ($d$= 0.4 mm). Center to center distance between two drops = 2.34 mm.

**5.2 Particle formation inside the droplet**

Hence, it is clear from the above discussion that, within a short period of time, almost uniform energy distribution is achieved due to fast energy transport and also the velocity and temperature distributions do not change with time once it is established, provided the disturbance introduced by particles on fluid motion is negligible. In fact, since nanoparticles are very small, effect of particles on the flow field can be neglected. Therefore, one can conveniently consider, an unit cell, considering a single droplet with surrounding fluid (fig. 1) and treat it as a batch reactor, where flow and temperature fields are already established (as obtained in section 5.1) and only temporal evolution of species- and particle size distribution is taking place (as dictated by already established velocity and temperature distributions). This approach eliminates the need of solving the more numerically complex, coupled problem and allows us to separate it into two sequential problems of lesser complexity. In this section, we discuss the variation of arithmetic mean diameter ($D_{10}$) of nanoparticles with the variation in different parameters.

### 5.2.1 Effect of Temperature

It is found in the experiment by Zukas and Gupta[55] that, as the system temperature increases, particle morphology changes from a plate like structure ($T=25$ °C) with length up to 500 nm, to a more spherical structure ($T= 40$ °C and above). However, in our theoretical analysis, we restrict to the prediction of only spherical morphologies, due to the limitation of our model being applicable to spherical structures only. The corresponding experimental results from Zukas and Gupta[55] have been shown in figure 4.

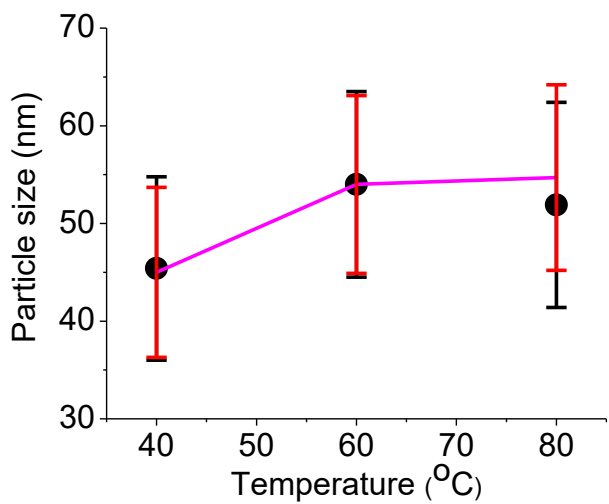

**Figure 4**: Effect of temperature on ZnO nanoparticle formation inside spherical droplets. Experimental conditions (Zukas and Gupta[55]): NaOH = 0.1M and Zinc acetate = 0.033 M**. The experimental data and simulation results represent mean and standard deviation of size distributions, not experimental error bars.** (•) represents the experimental mean and line (—) is used for connecting the simulation mean, as a guide to the eye. Standard deviation from PSD: Experiment (—) and simulation (—). Spherical drop diameter ($d$) =0.78 mm and channel diameter ($D_c$) =0.812 mm. Center to center distance between two drops = 2.34 mm.

Initially, mean particle size is observed to increase with increase in temperature from 40 °C to 60 °C. On further increase in temperature to 80 °C, particle size is found to decrease very slightly. However, polydispersity of particles shows a gradually increasing trend with increase in temperature, as shown by the range of the vertical lines against each data point, depicting standard deviation for each distribution. In this regard, it is to be noted that, Zukas and Gupta[55] did not provide any experimental error bar (sensitivity) on their mean diameter from repeated experimental runs. So we can see that, our simulation nicely captures this initial variation in both mean size and size distribution, with coagulation efficiency being a fitting parameter. To the credit of the model, the prediction of polydispersity is very good at all temperatures, justifying that the single parameter in the model is able to capture both mean and standard deviation of the experimental data at all temperatures. This further validates our model keeping synergy with the step by step validation of the model provided in the section S9. It is to be noted that, coagulation efficiency has been selected as the fitting parameter as it is the only unknown parameter in this model. There are ways to obtain the independent estimate of coagulation efficiency,[72-73] but that needs rigorous calculation. Therefore, we are estimating the coagulation efficiency by guessing a value of efficiency followed by comparing the resulting simulation results with the experimental data.

In general, effect of temperature is noticed through variation of reaction, nucleation and coagulation rates. In the present case, ZnO precipitation follows an instantaneous reaction kinetics. It is also observed in figure 2a that, even though there is a significant recirculation of liquids in both inside and outside the droplets, it is less pronounced towards the center of the microreactor, along its flow-axis. This indicates that, in the region close to the microreactor wall, reactants and products are transported mostly by convection, whereas in the central region, transport is still dominated by the diffusion mode. As temperature increases, diffusion transport of both reactants towards the interface region, where reaction is taking place, also increases. This enhances the rate of formation of ZnO, causing a fast build up of ZnO molecules and a high nucleation rate. This generates a large number of nuclei inside the droplet (eq. 9). In addition to that, the coagulation frequency function (eq. 9) also increases in direct proportion to temperature. Finally, increase in temperature also increases the surface-diffusion rate, resulting in increase of fusion rate of colliding particles through faster surface integration and consequently increasing coagulation efficiency, which is captured by the larger value of the fitted parameter $\beta$ (in eq. 9), as shown in table 1. Therefore, overall coagulation rate (eq. 9) increases due to these three effects, which results in the formation of particles of larger mean size with increase in temperature. The broadening of distribution with temperature may be attributed to the enhanced coagulation rate at higher temperature.

**Table 1:** Coagulation efficiency ($\beta$) as a function of temperature at
$$c_{Zn(OAc)_2} = 0.033M, c_{NaOH} = 0.1M$$

| T | $\beta$ |
|---|---|
| 40 | $1.3 \times 10^{-8}$ |
| 60 | $6.5 \times 10^{-8}$ |
| 80 | $1 \times 10^{-7}$ |

### 5.2.2 Effect of zinc acetate

Experimental results (Zukas and Gupta[55]) for the effect of zinc acetate on ZnO nanoparticle formation have been shown in figure 5.

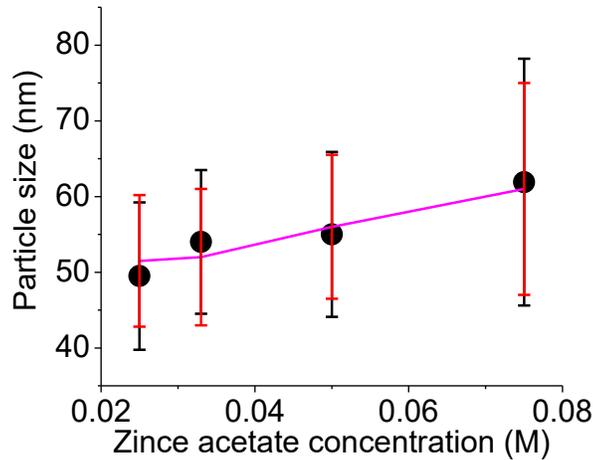

**Figure 5**: Effect of zinc acetate concentration on particle size and size distribution. Experimental conditions (Zukas and Gupta[55]): NaOH = 0.1 M and $T$= 60 °C. **Experimental data and simulation results represent mean and standard deviation of size distribution.** In the figure, (•) represents the experimental mean and the line (—) is used for connecting the simulation mean, as a guide to the eye. Standard deviation from PSD: Experiment (—) and simulation (—). Spherical drop diameter =0.78 mm and channel diameter =0.812 mm. Center to center distance between two drops = 2.34 mm.

In all cases, particles formed are having a spherical morphology. Mean particle size shows an increasing trend with increase in concentration of zinc acetate. Our model prediction gives reasonably good match with experimental data, with coagulation efficiency (β) being the only fitting parameter, which is given in table 2. Here we see that, coagulation efficiencies show significant variation with zinc acetate concentration, indicating its strong dependence on reactant concentration.

This can probably be explained by invoking the combined effect of mass transport towards the interface (as explained section 5.2.1) and increase in coagulation efficiency (due to increase in ionic strength). Since, stoichiometric amount of zinc acetate used in their experiments was lesser than that of NaOH in some cases, it can be concluded that the precipitation reaction to form ZnO growth units was limited by availability of former at the interface. Hence, it can be limited by transport of zinc acetate from bulk towards the interface. As the concentration increases, easy availability of zinc acetate at the interface region due to faster transport (as a result of higher concentration gradient) increases reaction rate, which subsequently increases nucleation rate. Therefore, as zinc acetate concentration increases, initial particle concentration also increases, causing the coagulation rate to become faster and finally bigger particles are formed with a larger size distribution. Although, increased number of particles can decrease the diffusion growth rate, its effect is insignificant. Moreover, since increased concentration of zinc acetate does not have any effect on solubility, OR rate remains the same.

**Table 2:** Coagulation efficiency as a function of zinc acetate concentration at $T = 60^{o}C, c_{NaOH} = 0.1M$

| Zinc acetate concentration (M) | β |
|---|---|
| 0.025 | $6.2 \times 10^{-8}$ |
| 0.033 | $6.5 \times 10^{-8}$ |
| 0.050 | $7.2 \times 10^{-8}$ |
| 0.075 | $6.2 \times 10^{-7}$ |

In case of very high concentration (0.075 M), stoichiometric amount of zinc acetate is higher than NaOH, and therefore, one can conclude that instantaneous reaction is not limited by zinc acetate. It may be possible that rate of formation of growth unit remains same after certain concentration of zinc acetate, but increased zinc acetate concentration leads to increase in coagulation rate due to decrease in double layer repulsion. In fact, this is observed by higher value of coagulation efficiency in table 2 and corresponding higher polydispersity in figure 5.

### 5.2.3a Effect of spherical droplet size

We have already seen in the previous hydrodynamic and thermal transport studies that, droplet size plays an important role in thermal energy transport. In order to study the effect of droplet size on the final particle size and size distribution, simulations have been carried out with spherical droplets of three different sizes [ratio of drop diameter ($d$) to channel diameter ($D_c$), $d/D_c$ =0.71, 0.81, 0.96] keeping all other parameters constant ($T$=60 °C, zinc acetate = 0.033M and NaOH = 0.1 M) and with the use of coagulation efficiency obtained in section 5.2.1 (temperature effect). Corresponding simulation results have been shown in figure 6. Since there is no report on effect of the droplet size over nanoparticle size and size distribution in the experimental investigation of Zukas and Gupta,[55] it has not been possible to compare our present simulation results with experimental data. In present simulation results, it is observed that, particle size increases with increase in droplet size, whereas polydispersity of particles shows a reverse trend. As the size of droplets decreases, convection-induced mixing of reactants as well as thermal energy transport both inside and outside the drops becomes less efficient, as observed by the absence of recirculation of fluid, in figure 2b, in case of a small drop. Due to the dipole structure of flow within the droplets, transport near the core region mainly takes place by diffusion. Same is true for surroundingexternal region too. In case of larger drops, presence of strong recirculation reduces the length scale over which the diffusion transport is the dominant mode. Therefore, in case of drops with relatively smaller size and resulting larger diffusion length, mass transfer limitation in both sides of the droplet interface is even more. This slows down rate of formation of growth units (ZnO), as less reactants are available at the reaction site. Consequently, less number of nuclei is formed. This eventually produces smaller size particles, due to lower diffusion growth rate and coagulation rate.

In a similar manner, as the heterogeneity in the distribution of reagents and thermal energy caused by inefficient mixing inside small drops increases, spatial variation in the growth rate of particles would also occur. This leads to the production of particles with a broader size distribution. On the contrary, as the droplets get bigger, mixing inside the droplets also becomes more pronounced, resulting in particles with low polydispersity and little large size. This small increase in particle size is not large as observed by a small value of coefficient of variation (CV) of21.3% in case of $d/D_c$ of 0.96, as compared to that of 33.4% in case of $d/D_c$ of 0.71.

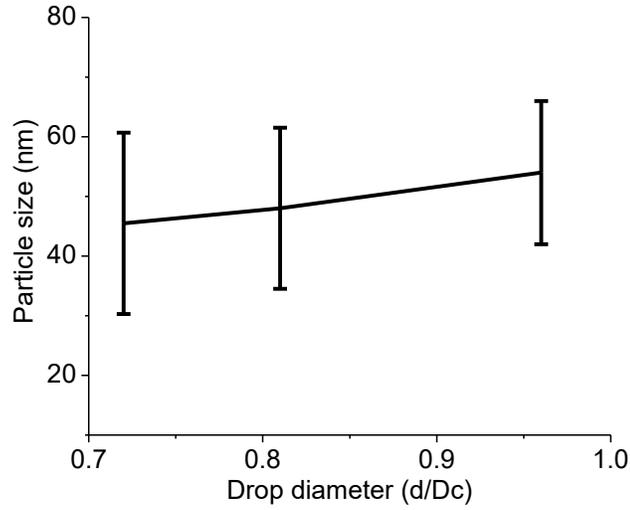

**Figure 6:** Effect of droplet size on particle size and size distribution. **Simulation results show mean and corresponding standard deviation.** Simulation conditions: $T$=60 °C, zinc acetate = 0.033M and NaOH = 0.1 M with $\beta = 6.5\times10^{-8}$ from temperature effect study. Channel diameter =0.812 mm. Ratio of spherical drop diameter ($d$) to channel diameter ($D_c$), $d/D_c$ =0.71, 0.81, 0.96. Center to center distance between two drops = 2.34 mm.

### 5.2.3b Selection of drop size

In section 5.2.3a, we have mentioned that, as the size of droplet decreases, mixing is limited by mass transport. It is worthy to mention here that, if the droplet size is decreased to a very small size, it may be possible that diffusion transport is enough to cause sufficient mixing of reactants within the droplets, even though circulation-induced mixing does not exist. In that case, heat transport limitation will come into play, as observed in figure 3b. Therefore, we restrict to relatively larger drops in the present investigation. In order to investigate this further, time scale analysis has been performed for different droplet sizes as shown in figure 7. Mass transport towards the interface occurs due to the combined effect of convective and diffusion mode. Time scale ($t_d$) for mass transport by diffusion has been calculated from Einstein's relation as follows

$$r^2 = 6D_m t_d \tag{12}$$

where, $r$ is the length scale over which diffusion takes place. Time scale for convective transport has been calculated from the corresponding vortex size and velocity within the droplet, obtained from our own simulation results of droplet-laden flow. Time scale for the energy transport is defined as the time required for the droplet center temperature to reach a value close to wall temperature. (see supplementary material section S6)

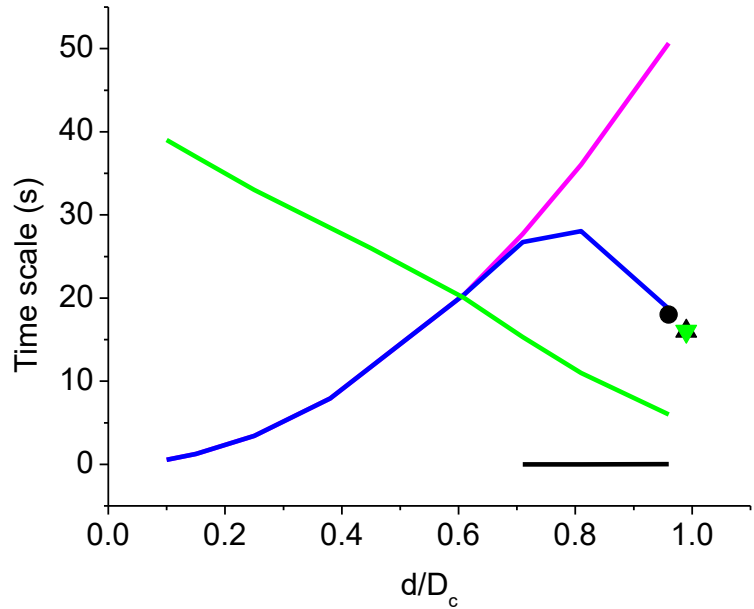

**Figure 7**: Time scale for thermal mixing (—). **Simulation in presence of fluid convection:** total time scale for mass transport (convection +diffusion) (---), time scale for mass transport by convection (—). **Simulation without convection (without hydrodynamics):** time scale for mass transport (by diffusion as mass transport occurs by diffusion mode only) (—); Earlier operating region ($d/D_c$) from literature: (•) Zukas and Gupta,[55] (▲) Khan et al.[3] and (▼) Khan et al.[70]

In figure 7, we observe that, total time scale for mass transport (adding contribution from both diffusion and liquid recirculation) from the center of droplet to the droplet interface initially increases with increase in drop size followed by a decrease. This has been explained by calculating individual diffusion and convective transport time scales as a function of droplet size (fig. 7). It shows that, convective transport is much faster compared to diffusion transport. For smaller drops, mass transport occurs solely by molecular diffusion. As droplet diameter increases, although convective transport inside drop increases, this increase is not pronounced, as length scale, over which molecular diffusion occurs, still keeps on increasing. Therefore, total time scale for mass transport increases with increase in drop size (figure 7). When droplet size becomes very close to the channel diameter, liquid recirculation inside droplet becomes more pronounced and this reduces the length scale over which diffusion transport takes place near the core of the droplet. Hence, we see a decrease in total time scale for mass transport (Vortices can occur in any length scales. It is to be noted that, all the calculations in this analysis is based on the largest vortex, approximate size of which is determined from the largest circular streamline inside the drop. It is not possible to calculate the exact size of the largest vortex in this particular system). If convection is switched off (by treating the system as a stagnant medium), we see that, time scale is continuously increasing with drop size as diffusion length scale is increasing. In contrast to time scale for mass transport, time scale for energy transport throughout shows a decreasing trend, with increase in droplet diameter, as in figure 7. This can be explained in a similar manner as in section 5.1, by invoking the effect of faster transport of hot fluid near the droplet interface from the channel wall, as drop size is increased. Therefore, even if smaller drops provide quick mixing by molecular diffusion alone,[1] it is not desirable for nanoparticle formation, because of thermal inhomogeneity. This design rule is also supported by earlier experimental reports on nanoparticle synthesis in segmented flow microfluidic reactor.[3,55,70] This has been shown in figure 7 by data points with green and black symbols, wherein droplet size is maintained close to their respective channel diameter, as concluded to be beneficial through our present simulation outcome.

**5.2.4 Effect of non-spherical droplets**

Here, we use droplets of non-spherical shape (obtained by elongating only the major axis ($L$) of an eventual ellipse-like drop-shape, with minor axis being kept same as the spherical drop), in order to study its effect on the final PSD. In order to achieve this, we perform simulation with droplets of three different major axis length ($L/D_c$: 0.96, 1.5 and 2; $D_c$= 0.812 mm), but of same minor axis length ($d$=0.78 mm). Corresponding simulation results are shown in figure 8 with mean size and size distribution. We see that, particle size increases insignificantly with increase in degree of non-sphericity of droplet, while size distribution shows significant decrease in polydispersity. Similar to the spherical drops, one can easily explain the present trend by invoking the mixing effect. In case of non-spherical drops, better mixing is achieved as degree of non-sphericity of the drops increases as is observed in figure 2c. This reduces spatial differential growth of particles and thereby reduces the polydispersity (fig. 8). The effect of faster transport is also observed by relatively higher conversion of reactants and bigger sized particles.

When the present simulation results are compared with that of a spherical droplet ($L/D_c =$ 0.96), particles are in general bigger with lower polydispersity in the present case. This is because convection-induced mixing is much stronger (figure 2c) inside as well as outside the non-spherical droplets as compared to that of spherical droplets. This increases transport rate of reagents as well as reduces the heterogeneity in temperature, species and particle number density inside the non-spherical droplets, causing bigger particles with lower polydispersity. For the non-spherical drop with highest non-sphericity ($L/D_c=2$; fig. 8), mixing is so efficient that, polydispersity (reduced by a factor of ~2) is strikingly low. Even though particle size is increased (in an insignificant extent) with increase in degree of non-sphericity (fig. 8), present finding seems to be quite promising to many industrial applications, as the variability of distribution (11% and 21.3% for $L/D_c$ of 2 and 0.96, respectively) with respect to mean is quite low in case of droplet with highest non-sphericity. Further, the synthesis method, proposed herein, has a great implication, if scale up aspects of droplet based reactor is considered for large scale production of ZnO. This is further supported by reasonable conversion (~91%) of reactants within non-spherical drops of $L/D_c$ of 2, compared to that of ~95% in case of spherical drops of $d/D_c$ of 0.96.

In order to understand, whether particles with bigger size are simply produced due to increase in drop volume (i.e. mass of reagents), simulation has been carried out (fig. 8) in absence of fluid convection (treating droplet and its surrounding as a stagnant medium) in all three case of $L/D_c$, keeping all other conditions same. This shows that, particles obtained in this case are relatively smaller with very high polydispersity (CV of 32% for $L/D_c$ of 2 and without convection). This supports the already observed effect of convection-induced mixing inside and outside the droplets being the reason behind the making of larger particles with lower polydispersity (with CV of 11% for $L/D_c=2$ and with convection). Once the initial conversion of reactants is over near the interface region, some of the growth units produced will diffuse towards the core of the droplet and will nucleate. Since diffusion transport of reactants is very slow, next encounter of reactants also gets delayed and this will result in less number of growth units and also fewer nucleuses, as compared to that observed in presence of convection. This results in smaller size particles. With time, the reaction front moves very slowly towards the core of droplet and same sequence of events occur as in the initial phase. This introduces significant differential growth of nanoparticles at different spatial positions and consequently polydisperse particles are produced. Since reaction front moves very slowly towards the core in absence of convection, conversion is also relatively low (67%), as compared to that in presence of convection.

In order to further understand that, how the presence of reactants in separate phases affect particle formation and thereby the final PSD, simulation has been performed in continuous single phase laminar flow microreactor (figure S5 in the supporting material section S6) too. This has been done by keeping all the conditions same as in droplet based microreactor, except that reactants are present in a single phase and are assumed to be well mixed from the beginning itself. Calculations show that, particles produced in the single phase case are relatively bigger in size (75 nm) with a broader size distribution (with standard deviation ~30 nm and CV =40%) as compared to CV of only 11% in case of $L/D_c$ of 2 in droplet based microreactor, with convection. This is because, in case of laminar flow reactor, all of the growth units are available from the beginning (because of the well mixed assumption and instantaneous reaction) i.e. at the entrance to the reactor. This results in an extremely large numbers of nuclei (due to extremely high supersaturation). As a consequence of this, coagulation rate (which is driven by both Brownian motion and shear-induced coagulation in laminar flow reactor) is also extremely high, resulting in bigger particles with a broader size distribution. As the particle size increases, shear induced coagualtion rate also increases, further increasing the polydispersity. This has also been observed by Khan *et al.*[3] and many others,[20-25] in their laminar flow single phase synthesis, confirming that, presence of shear coagulation produces highly polydisperse particles.

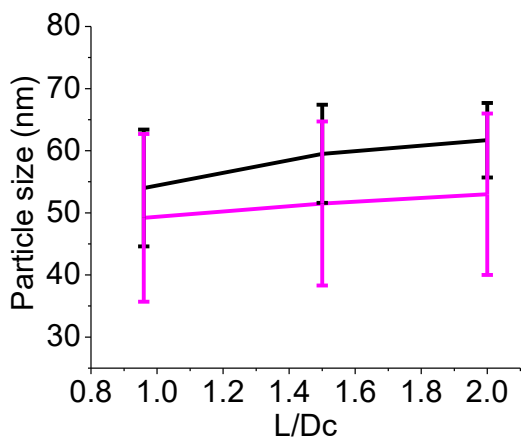

**Figure 8:** Effect of degree of non-sphericity of droplet on particle size and size distribution. Simulation conditions: $T$=60 °C, zinc acetate = 0.033 M and NaOH = 0.1 M with $\beta$ = 6.5×10$^{-8}$ from temperature effect study. Channel diameter ($D_c$) =0.812 mm. Non-spherical drop minor

axis length = 0.78 mm. Non-spherical drop major axis length ($L$) = 0.78, 1.22 and 1.62 nm. Center to center distance between two elongated ellipsoidal drops = 2.34 mm. Simulation curves: with convection-induced mixing (──) and without convection effect (──). **Simulation results show mean and corresponding standard deviation.**

## 6 Conclusions

For the first time through this work, we have theoretically investigated the mechanistic details of nanoparticle formation in an interphase droplet based microfluidic synthesis. This will facilitate the design of corresponding microfluidic devices and experiments therein, in order to produce particles of smaller mean size with lower polydispersity, which is the principle objective of current research. An experimentally validated model has been proposed, by coupling the effect of fluid dynamics and thermal transport with the particle formation mechanism. We find that, the final mean particle size and size distribution are strongly affected by the droplet dynamics induced transport and distribution of reagents and thermal energy inside and outside the droplets.

Our present analysis also shows that, droplet size and shape are crucial parameters in controlling the final PSD, specially the polydispersity. It is found that particles with lower polydispersity can be synthesized by increasing the droplet size, which facilitates the internal circulation of fluid, but with very little increase in particle size. Polydispersity can be further reduced to a great extent by introducing non-spherical drops of larger volume. Though mean particle size increases with increase in size of the spherical droplet and degree of non-sphericity of non-spherical drops, the increase is not significant, when change in polydispersity is considered. Although there is very little work on the formation of non-spherical drops and synthesis of nanoparticles therein, present findings can be applicable to plug shaped droplets too, as latter also provides better mixing, as compared to a spherical drop.

Further analysis shows that, even though smaller drops provides reasonably good mixing by molecular diffusion itself (eliminating the need of convection-induced mixing), energy transport limitation limits the suitability of smaller drops as a host for nanoparticle formation. This suggests the importance of mixing, in a segmented flow microreactor synthesis of nanoparticles. In fact, our finding shows that, maintaining a certain operational regime (i.e. ratio of droplet size to microreactor channel diameter > 0.8) is crucial, in obtaining particles with lower PSD (CV of 21.3% for $d/D_c = 0.96$); without which particles with higher PSD (with CV of 33% for $d/D_c = 0.71$) are obtained.

Thus we find that, multiphase synthesis with reactants initially present in separate phases produces relatively smaller particles with significantly lower polydispersity, as compared to that in a single phase synthesis []. Presence of reagents in separate phases acts, as if it is releasing the material in a controlled manner, thereby controlling the particle formation step. Therefore, by suitably controlling the design and operating parameters, one can fine-tune the particle size and its distribution, in an efficient manner, which is not possible in any other microfluidic synthesis method. This shows the relevance of the droplet based interphase synthesis. Though the present analysis considers the synthesis of ZnO nanoparticles, it can be extended for industrial scale synthesis of other materials like $Fe_2O_3$, Ag, $BaSO_4$, CdS, $CeO_2$, $TiO_2$ etc. [71-76]. This is because nanoparticles with low polydispersity have immense applications in a variety of fields including catalysis, drug delivery, biosensing, magnetic resonance imaging, storage devices like fuel cells, batteries etc., and antimicrobial agents [92-95]. Apart from the importance in the area of synthesis of nanoparticles with low polydispersity, findings from the present work will be useful in the area of industrial scale separation, extraction, and purification of inorganic and biological materials like platelates, tumar and cancer cells, proteins, inorganic microparticles etc. [77-87]. Finding from this analysis is also having immense significance in the area of cooling of microelectronic devices, boiling in a microchannel, and in multiphase flow reactors in industry [88-91].

Finally, we point out the uniqueness of the present work in several aspects, which have not been explored in any of the previous works on nanoparticle formation:

- To the best of our knowledge, for the first time, it explores the physics of nanoparticle formation in any segmented flow microreactor. This has been possible by including the effect of fluid and thermal dynamics on particle evolution. This makes it possible to apply this modelling approach and its simulation framework directly to any other multiphase flow synthesis of nanoparticles or crystallization of inorganic or biological materials in any kind of reactor, as it incorporates the effect of hydrodynamics of multiphase flow.

- To the best of our knowledge, this is the first time that, particle formation has been studied, considering the spatial variation of reactants in both continuous and dispersed phase, in a multiphase system, where reactants are initially present in separate phases. It has been experimentally found that droplet based interphase synthesis of nanoparticle offers better control over reactant addition and thereby on nanoparticle size distribution, compared to other multiphase synthesis. Therefore understanding from present theoretical analysis will help in designing the interphase synthesis, involving other reactors too.

- The present work explicitly studies the effect of hydrodynamics on the mass transport dynamics inside and outside the droplet, in a segmented flow reactor. This will further help in understanding the physics involved with various processes, such as extraction, separation, crystallization and synthesis of chemicals etc, even more in a segmented flow multiphase system. We find that there exists a critical ratio of droplet diameter to channel diameter, above which time scale for mass transport decreases and operation of reactor above this value provides particles with desired characteristics.

- To the best of our knowledge, this work for the first time presents the effect of droplet size and shape on the mass transport dynamics and the particle evolution dynamics therein. We show that, droplets with bigger size produces particles with low polydispersity, which is further reduced with non-spherical drops with relatively larger volume, maintaining approximately same conversion. This has great implication if scale up aspect of the reactor is considered for large scale industrial production of nanomaterials is concerned.

**Acknowledgements**

Authors acknowledge Prof. Nivedita Gupta and Mr. B. G. Zukas, Department of Chemical Engineering, New Hampshire, USA, for bringing this droplet based synthesis to our attention, and for useful research discussion on their experimental data and valuable feedback.

**Supplementary material**

A schematic of experimental set up and a detailed discussion on nucleation, diffusion growth, coagulation, Ostwald Ripening and experimental procedure have been presented in the supplementary material section available online.

**Nomenclature**

| | |
|---|---|
| $A_{ll}$ | surface area of liquid-liquid interface (m$^2$) |
| $A$ | pre-exponential coefficient in classical nucleation kernel (number/m$^3$s) |
| $c_r$ | equilibrium concentration of ZnO at the surface of particle of radius $r$ (mole/m$^3$) |
| $c_{ZnO}$ | bulk concentration of ZnO (mole/m$^3$) |
| $c_{ZnO,s}$ | solubility of ZnO (mole/m$^3$) |
| $d$ | diameter of droplet (m) |
| $D_c$ | microreactor channel diameter (m) |
| $D_m$ | molecular diffusion co-efficient (m$^2$/s) |
| $E_m$ | mass averaged energy (J/kg) |
| $E^q$ | $q^{th}$ phase energy (J/kg) |
| $F_{sv}$ | volume force (N/m$^3$) |
| $F_{sa}$ | continuum surface force (N/m$^2$) |
| $G_v$ | volumetric growth rate of particle (m$^3$/s) |
| $\hat{j}_i^p$ | diffusion flux of species $i$ in the primary phase (kg/m$^2$s) |
| $\hat{j}_i^q$ | diffusion flux of species $i$ in the phase, $q$ (kg/m$^2$s) |
| $k_B$ | Boltzmann constant (J/K) |
| $k_{eff}$ | effective thermal conductivity (W/mK) |
| $L$ | length of major axis of non-spherical drops (m) |
| $\dot{m}^{ps}$ | mass transfer rate of species $i$ from primary to secondary phase (kg/m$^3$s) |
| $\dot{m}^{sp}$ | mass transfer rate of species $i$ from secondary to primary phase (kg/m$^3$s) |
| $\dot{m}^{ts}$ | mass transfer rate from tertiary to secondary phase (kg/m$^3$s) |
| $\dot{m}^{st}$ | mass transfer rate from secondary to tertiary phase (kg/m$^3$s) |
| $n$ | number density function of particles (number/m$^6$) |
| $n(v,0)$ | particle number density in the system initially (number/m$^6$) |

| $n(0,t)$ | particle number density in zeroth bin at time $t$ (number/m$^6$) |
| --- | --- |
| $n_0$ | nucleation rate (number/m$^3$s) |
| $\hat{n}$ | unit norml vector |
| $\overline{N}_i$ | total flux of $i^{\text{th}}$ species in the corresponding phase (kg/m$^2$s) |
| $p$ | primary phase |
| $P_r$ | pressure |
| $q_B$ | Brownian coagualtion frequency function (m$^3$/s) |
| $r$ | particle radius variable (m) |
| $r_{cn}$ | size of nucleus (m) |
| $R$ | Universal gas constant |
| $Re_d$ | droplet Reynolds number |
| $Re_b$ | bulk Reynolds number |
| $R_i^q$ | loss or gain of species $i$ due to reaction in the $q^{\text{th}}$ phase (kg/m$^3$s) |
| $s$ | secondary phase |
| $S$ | super saturation of the ZnO in the secondary phase |
| $S_h$ | volumetric source term (J/m$^3$s) |
| $T$ | temperature (K) |
| $T_w$ | constant temperature maintained at reactor walls (K) |
| $T_i$ | initial temperature (K) |
| $T_m$ | mass averaged temperature (K) |
| $\overline{u}$ | combined phase velocity vector (m/s) |
| $\overline{u}^q$ | $q^{\text{th}}$ phase velocity vector (m/s) |
| $\overline{u}^p$ | primary phase velocity vector (m/s) |
| $\overline{u}^s$ | secondary phase velocity vector (m/s) |
| $v, v'$ | particle volume variable (m$^3$) |
| $v_i, v_j$ | volumes of colliding particles (m$^3$) |
| $v_m$ | molecular volume of ZnO (m$^3$) |
| $V_m$ | molar volume of ZnO (m$^3$) |
| $x$ | $x$ direction in Cartesian co-ordinate |
| $y$ | $y$ direction in Cartesian co-ordinate |
| $Y_{i,0}$ | initial mass fraction of $i^{th}$ species in the corresponding phase |
| $Y_i$ | mass fraction of species $i$ in the corresponding phase |

**Greek letters**

| | |
|---|---|
| $\alpha^q$ | $q^{th}$ phase (primary or secondary phase) volume fraction |
| $\alpha^p$ | primary phase volume fraction |
| $\alpha^s$ | secondary phase volume fraction |
| $\beta$ | Brownian coagulation efficiency |
| $\kappa_c$ | curvature of liquid-liquid interface |
| $\mu$ | viscosity of medium (kg/ms) |
| $v$ | kinematic viscosity (m$^2$/s) |
| $\rho_m$ | combined phase density (kg/m$^3$) |
| $\rho^q$ | $q^{th}$ phase (primary or secondary phase) density (kg/m$^3$) |
| $\rho^p$ | primary phase density (kg/m$^3$) |
| $\rho^s$ | secondary phase density (kg/m$^3$) |
| $\sigma$ | interfacial tension of ZnO-water system (N/m) |
| $\sigma_{ll}$ | surface tension of liquid-liquid interface (N/m) |
| $\nabla_s$ | surface tangential gradient operator |
| $\nabla_v$ | operator in particle volume co-ordinate (internal co-ordinate) |
| $\bar{\bar{\tau}}$ | shear stress tensor in VOF model |

### 75. Manufacturing nanomaterials: from research to industry

Manufacturing Rev.
**Volume** 1, 2014

**Article Number** 11
**Number of page(s)** 19

**Costas A. Charitidis**[*], **Pantelitsa Georgiou, Malamatenia A. Koklioti, Aikaterini-Flora Trompeta and Vasileios Markakis**

### 72. Preparation of Silver Nanoparticles and Their Industrial and Biomedical Applications: A Comprehensive Review

Adnan Haider and Inn-Kyu Kang, Advances in Materials Science and Engineering
Volume 2015, Article ID 165257, 16 pages

82. # The use of microfluidic devices in solvent extraction

Davide Ciceri

Jilska M. Perera

Geoffrey W. Stevens

Volume89, Issue6, June 2014

Pages 771-786

83. ## Microfluidic extraction, stretching and analysis of human chromosomal DNA from single cells†

Jaime J. Benítez,[a,‡] Juraj Topolancik,[a,‡] Harvey C. Tian,[d] Christopher B. Wallin,[a] David R. Latulippe,[a] Kylan Szeto,[a] Patrick J. Murphy,[b] Benjamin R. Cipriany,[e] Stephen L. Levy,[c] Paul D. Soloway,[b] and Harold G. Craighead, Lab Chip. 2012 Nov 21; 12(22): 4848–4854.

**84. Continuous separation of microparticles in a microfluidic channel *via* the elasto-inertial effect of non-Newtonian fluid†**

Jeonghun Nam,[‡a] Hyunjung Lim,[‡a] Dookon Kim,[a] Hyunwook Jung[b] and Sehyun Shin[*a], Issue 7, 2012

Lab on a chip.